\documentclass[12pt,twoside,A4]{article}
\usepackage{thumbpdf,lmodern,bm}
\usepackage{graphicx,subcaption}
\usepackage{amsmath} 	        
\usepackage{amssymb}
\usepackage[table, dvipsnames]{xcolor}
\usepackage{apacite}
\usepackage{natbib}

\usepackage{hyperref}
\hypersetup{pdfstartview={XYZ null null 1.00}}

\begin{document}

\title{LNIRT: An \textit{R} Package for Joint Modeling of Response Accuracy and Times}

\author{Jean-Paul Fox;\footnote{University of Twente, The Netherlands} Konrad Klotzke \footnote{University of Twente, The Netherlands}; Ahmet Salih Simsek \footnote{University of Kirsehir Ahi Evran, Turkey}}

\date{}

\maketitle
\begin{abstract}
In \textit{computer-based testing} it has become standard to collect response accuracy (RA) and response times (RTs) for each test item. IRT models are used to measure a latent variable (e.g., ability, intelligence) using the RA observations. The information in the RTs can help to improve routine operations in (educational) testing, and provide information about speed of working. In modern applications, the joint models are needed to integrate RT information in a test analysis. The R-package LNIRT supports fitting joint models through a user-friendly setup which only requires specifying RA, RT data, and the total number of Gibbs sampling iterations. More detailed specifications of the analysis are optional. The main results can be reported through the summary functions, but output can also be analysed with Markov chain Monte Carlo (MCMC) output tools (i.e., coda, mcmcse). The main functionality of the LNIRT package is illustrated with two real data applications.

\noindent\textbf{keywords:} item-response theory, response times, joint modeling, R-Package.
\end{abstract}

\section{Introduction} \label{sec:intro}
When a computer-based test is administered, next to response accuracy (RA) response times (RTs) can be automatically recorded. The information in the RTs can help to improve routine operations in testing, such as item calibration, adaptive item selection, latent ability estimation, as well as to explore and measure factors that influence the performances on the test.

In the literature, the modelling of RTs has been approached from three different angles. One approach is to add time parameters to an item response theory (IRT) model \citep[see, e.g., ][]{Roskam97, Thissen83, Verhelst97}. A second approach is to model the RTs separately from the responses \citep{Maris93, Scheiblechner79}. In a third approach, introduced by \cite{Linden2007}, the RTs and RA are modeled hierarchically. At the first level, both the distributions of RA and RTs are assumed to follow separate models, each with a different set of person and item parameters. The person parameters represent the speed and accuracy (or ability) of the test taker on the items. A test taker's choice of speed and accuracy is generally constrained by a tradeoff, where accuracy can be improved by working slower. At this first level of modeling, the RTs and RA can be assumed to be conditionally independently distributed given the speed and accuracy parameters, respectively. However, at the second level, these parameters are allowed to be dependent. This leads to a hierarchical modeling framework in which the relation between speed and accuracy is defined at a higher level of modeling.

RTs have a natural lower bound at zero. Therefore, a log-normal distribution is used to model the RTs, and subsequently, the logarithm of the RTs (log RTs) are assumed to be normally distributed. The choice of a lognormal distribution is a classic one in RT research. For RTs on test items, this assumption was made earlier by, for example, \cite{Thissen83}, \cite{Schnipke97a}, and \cite{Linden99}. Each of these studies showed a good fit of a lognormal distribution. Both the binomial distribution of RA and the normal distribution of the log RTs can be given a traditional IRT parameterization. The binomial parameter for RA has the structure of the two-parameter normal-ogive model \citep{Lord68}. The distribution of the RTs has a parameterization close to that of an IRT model for continuous response data \citep[see, e.g., ][]{Samejima73, Shi98}. RA and RTs are conditionally independently distributed. Their joint distribution is the product of a binomial and a normal distribution. This defines the level-1 model of the joint model for the analysis of RTs and RA for measuring test takers's speed and ability on test items, respectively. The joint model supports the use of available collateral information about each of the parameters: The RTs serve as collateral information that is used to estimate the parameters of the IRT model. Conversely, RA is used as collateral information when estimating the parameters of the RT model \citep{LindenCol2010}.

The {R}-package {LNIRT} \citep{LNIRT} supports the Bayesian joint modeling of RA and RTs and comprehends several Gibbs samplers for parameter estimation. The R-package is available on CRAN: \url{https://cran.r-project.org/web/packages/LNIRT/index.html}. The R-package {LNIRT} is the successor of the {cirt} package of \citet{fox2007}. The {cirt} program was implemented in Visual Pro FORTRAN, which led to problems in maintaining and updating the software. The program {LNIRT} has been developed in {R}, which makes it open source and makes the maintenance of the program much easier. The Gibbs sampler for the hierarchical joint model implemented in the {cirt} package has been integrated in the {LNIRT} package, and several important extensions have been added to LNIRT (see points 2-5 below). The {LNIRT} package offers several important contributions for jointly analysing item patterns of RA and RTs:
\begin{enumerate}
\item Statistical computations are done through a powerful Gibbs sampling method that allows for fast MCMC convergence and efficient MCMC sampling.

\item Extensive residual analysis tools are available for the evaluation of item- and person fit, and outlier detection \citep{Fox2017person}.

\item Different parameterizations of the joint model can be applied. For instance, the log-normal RT model can be fitted with the parameterization of \cite{Linden2007} or of \cite{kleinentink2008}.

\item Explanatory variables can be included at the item and person level, and missing item response data can be treated as missing at random (MAR) or missing by design (i.e. incomplete test design)\citep{van2016joint}.

\item It offers a generalized measurement model for RTs in which working speed can be modeled through a latent growth component to allow for differential working speed across items \citep{Fox2016joint}.
\end{enumerate}

The remainder of this paper discusses the main contributions of the package. Two detailed data examples are presented which show how the package can be used. In Section \ref{sec:models} the joint model is described, which includes the different measurement model parameterizations, the inclusion of explanatory variables at the item and person level, and (hyper) prior distributions. In Section \ref{sec:modelfit}, statistical tools are discussed to evaluate the fit of the model. This includes person-fit and item-fit tools to flag extreme persons and items under the model, respectively. Section \ref{sec:application} describes an illustration of the package by analysing the Credential data set collected from 1636 candidates who took the licensure exam \citep{cizek2016handbook}. Furthermore, the joint model with variable working speed is discussed using the Amsterdam chess data \citep{van2005psychometric}. Then, in Section \ref{sec:summary}, the conclusions are given.

\section{The Joint Model} \label{sec:models}

A hierarchical modeling procedure is followed. At level 1, separate measurement models are defined for the RA and RTs. At level 2, a distributional structure is defined for the level-1 parameters. Subsequently, hyper-prior distributions are specified for the prior parameters.

\subsection{Level 1: Measurement Models} \label{sec:level1}

\subsubsection{RA Model}
Item responses to a set of items indexed $k=1,\dots,K$ are taken to be stored in an $N$ by $K$ matrix $\mathbf{Y}$. The response patterns are
characterized by both the test takers and the items. A two-parameter IRT model defines the probability of a correct response given person and item parameters \citep[see, e.g.,][]{Lord68}. Let $\theta_{i}$ denote the ability of test taker $i$ $(i=1,\ldots,N)$. Then, the probability of a correct response to item $k$ is defined as:
\begin{equation} \label{eq:responsemodel}
P(Y_{ik}=1\mid \theta _{i},a_{k},b_{k})=\Phi (a_{k}\theta _{i}-b_{k}),
\end{equation}
where $a_{k}$ and $b_{k}$ are generally known as the discrimination
parameter and difficulty parameter of item $k$, respectively, and $\Phi$ denotes the normal cumulative distribution function. When defining the item difficulties on the same scale as the ability scale, additional brackets need to be placed in the mean component. Then, the probability of a correct response to item $k$ is given by,
\begin{equation} \label{eq:responsemodel1}
P\left(Y_{ik}=1\mid \theta _{i},a_{k},b_{k}\right)=\Phi\left(a_{k}\left(\theta _{i}-\tilde{b}_{k}\right)\right),
\end{equation}
where $\theta _{i}$ and $\tilde{b}_{k}$ are defined on the same scale. In {LNIRT}, both parameterizations are implemented. Note that the item difficulty parameters in Equation (\ref{eq:responsemodel}) and Equation (\ref{eq:responsemodel1}) are not directly comparable, and are defined on different scales. For the three-parameter model, a guessing parameter $c_k$ is introduced, representing the probability of guessing item $k$ correctly, and this leads to the following measurement model for the success probability:
\begin{equation} \label{eq:responsemodel2}
P(Y_{ik}=1\mid \theta _{i},a_{k},b_{k},c_k)= c_k + (1-c_k)\Phi (a_{k}\theta _{i}-b_{k}).
\end{equation}

\subsubsection{RT Model}
The log RTs are stored in an $N\times K$ matrix $\mathbf{RT}$. When assuming a constant working speed, each test taker works with a constant speed represented by $\zeta_{i}$. The time needed to complete an item also depends on item characteristic parameters. They are denoted as $\phi _{k}$ and $\lambda _{k}$, and can be seen as a time-discrimination and time-intensity parameter, respectively. The logarithm of the RTs, $RT_{ik}$, are assumed to be normally distributed:
\begin{eqnarray} \label{eq:responsetimemodel}
RT_{ik} & = & \lambda_k - \phi_k \zeta_i +\epsilon_{ik}  \\
\epsilon_{ik} & \sim & N\left(0,\sigma^2_{\epsilon_k} \right), \nonumber
\end{eqnarray}
where the time-intensity parameter $\lambda_k$ represents the average time needed to complete the item (on a logarithmic scale), the speed parameter, $\zeta_i$, represents the working speed of test taker $i$, and the time-discrimination parameter, $\phi_k$, the item-specific effect of working speed on the $RT$. Increasing the time-intensity $\lambda _{k}$ leads to a positive shift of the location of the time distribution on the item. Likewise, an increase in the speed parameter $\zeta _{i}$ leads to a negative shift. In the same way as for the item response model in Equation (\ref{eq:responsemodel1}), the time intensities can be defined on the same scale as the speed parameter. It follows that
\begin{eqnarray} \label{eq:responsetimemodel1}
RT_{ik} & = & \phi_k \left(\tilde{\lambda}_k - \zeta_i\right) +\epsilon_{ik},
\end{eqnarray}
where the time-discrimination parameter operates on the term $\tilde{\lambda}_k - \zeta_i$. The time-discrimination parameter was introduced as a slope parameter for speed, which models the sensitivity of the item for different speed-levels of the test takers \citep{fox2007,kleinentink2008}. This time-discrimination parameter differs from the time-discrimination parameter defined by \cite{Linden2007}. In his approach, the reciprocal of the standard deviation of the measurement error is defined to be the time discrimination. The time discriminations in Equation (\ref{eq:responsetimemodel}) also model covariances between RTs \citep{Fox2016joint}, which makes the model more flexible than the model of \cite{Linden2007}. Furthermore, the additional error term in Equation (\ref{eq:responsetimemodel}) can model variations in RTs due to stochastic behavior of a test taker. When test takers operate with different speed values, take small pauses during the test, or change their time management, the RTs might show more systematic variation than explained by the structural mean term. The item-specific error component might accommodate for these differences and avoids bias in the parameter estimates.

An implicit assumption of the RT model is that working speed is constant during the test. This means that, whatever the conditions under which the test is taken, the test takers are assumed to settle on a level of speed at the beginning of the test and then stick to it. In Section \ref{sec:diffworkingspeed}, the differential working-speed model is discussed to model changes in working speed, for example, due to fatigue or the adoption of a new strategy during the test.

\subsection{Level 2 and 3: Structural Person and Item Parameter Models} \label{sec:level2}
\subsubsection{Persons}
A bivariate normal (population) distribution is defined for the ability and speed parameter,
\begin{eqnarray}
(\theta_i ,\zeta_i) &\sim& \mathcal{N}_{2}(\mathbf{\mu}_{P},\mathbf{\Sigma}_{P}) \label{priorperson}\\
\mathbf{\mu }_{P} &=& (\mu _{\theta },\mu _{\zeta }) \nonumber \\
\mathbf{\Sigma }_{P} &=& \left(
\begin{tabular}{cc}
$\sigma_{\theta }^{2}$ & $\rho$ \\
$\rho$ & $\sigma_{\zeta }^{2}$
\end{tabular}
\right). \nonumber
\end{eqnarray}
The covariance between the person parameters is represented by $\rho$. The level-2 model for speed and ability can be considered to represent a population distribution for the test takers. The test takers are defined to be exchangeable, and the distribution represents the prior for the person parameters. Without an identification restriction(s) on the variance parameters, the hyperprior for the covariance matrix $\mathbf{\Sigma}_{P}$ is the inverse-Wishart distribution with degrees of freedom $\nu_{P}$ and scale parameter $V_{P}$.

\subsubsection{Items}
A multivariate normal distribution is specified for the item parameters,
\begin{eqnarray} \label{prioritem}
(a_{k},b_{k},\phi _{k},\lambda _{k}) &\sim& \mathcal{N}_4(\mathbf{\mu
}_{I},\mathbf{\Sigma }_{I}), \\
\mathbf{\Sigma }_{I} &=& \left(
\begin{array}{cc}
\mathbf{\Sigma }_{a,b} & \mathbf{\Sigma }_{(a,b),(\phi ,\lambda )} \\
\mathbf{\Sigma }_{(\phi ,\lambda ),(a,b)} & \mathbf{\Sigma }_{\phi
,\lambda }
\end{array} \right)  \nonumber \\
&=&\left(
\begin{array}{cccc}
\sigma _{a}^2 & \sigma _{a,b} & \sigma _{a,\phi } & \sigma _{a,\lambda } \\
\sigma _{b,a} & \sigma _{b}^2 & \sigma _{b,\phi } & \sigma _{b,\lambda } \\
\sigma _{\phi ,a} & \sigma _{\phi ,b} & \sigma _{\phi }^2 & \sigma _{\phi
,\lambda } \\
\sigma _{\lambda ,a} & \sigma _{\lambda ,b} & \sigma _{\lambda ,\phi } &
\sigma _{\lambda }^2
\end{array} \right). \nonumber
\end{eqnarray}
Parameters $a_k$ and $\phi_k$ are restricted to be positive. The covariance matrix for the item parameters allows for correlation between item parameters. The hyperprior for $(\mathbf{\mu}_{I},\mathbf{\Sigma }_{I})$ is a normal-inverse-Wishart distribution:
\begin{eqnarray}
\mathbf{\Sigma }_{I} &\sim & IW_{\nu _{I}}\left(V_{I}^{-1}\right)
\\
\mathbf{\mu }_{I}\mid \mathbf{\Sigma }_{I} &\sim &\mathcal{N}\left(%
\mathbf{\mu }_{0},\mathbf{\Sigma }_{I}/\kappa \right),
\end{eqnarray}
where $\nu_{I}$ and $V_{I}$ are the degrees of freedom and scale matrix of the inverse Wishart distribution, $\mathbf{\mu}_{0}$ is the prior mean and $\kappa$ the number of prior measurements, which is given a default value of one to specify a vaguely informative prior. A Beta prior is specified for the guessing parameter, $c_k$, where the default hyper-parameter values are 20 and 80.  This leads to a prior proportion of guessing of 1/5 with a standard deviation of .04.

\subsubsection{Explanatory Variables}
The multivariate models for persons and item parameters can be extended to include explanatory variables. Let $\mathbf{X}_{\theta}$  denote the predictors for the ability parameter and $\mathbf{X}_{\zeta}$ for the speed parameter. The mean component for the person parameters can be expressed as
\begin{eqnarray}
\mu_{\theta} = \mathbf{X}_{\theta}\mathbf{\beta}_{\theta}, \, \mu_{\zeta} = \mathbf{X}_{\zeta}\mathbf{\beta}_{\zeta}. \nonumber
\end{eqnarray}
For the mean component of the difficulty and time-intensity parameters a similar extension is defined,
\begin{eqnarray}
\mu_{b} = \mathbf{X}_{b}\mathbf{\beta}_{b}, \, \mu_{\lambda} = \mathbf{X}_{\lambda}\mathbf{\beta}_{\lambda}. \nonumber
\end{eqnarray}
Explanatory information can be included to explain differences between persons and item characteristics. Noninformative normal priors are defined for the regression parameters with a mean of zero and a large variance. In the {LNIRT} package, the option to include predictors for discrimination and time discrimination are not implemented, since the variance in parameter values is (very) small. For categorical predictor variables, dummy coding is required.

\subsection{Parameter Estimation and Model Identification} \label{sec:estimation}
IRT models are usually identified by fixing the mean and variance of the latent variable to zero and one, respectively. Typically, this can be done directly by restricting the prior mean $\mu_{\theta}$ and variance $\sigma^2_{\theta}$, or by putting restrictions on the item parameters. The joint model can be identified in the same way. However, for the identification rules in the package {LNIRT} restricting the variance of a person parameter has been avoided. When restricting the variance of a (random) person parameter, the covariance matrix in Equation (\ref{priorperson}) is also restricted, and the inverse-Wishart distribution does not apply to a restricted covariance matrix. For this scenario, \citet{kleinentink2008} redefined the prior for the person parameters, where $\rho$ becomes a regression parameter in the regression of working speed on ability with the working speed variance included in the error variance. However, this identification procedure would also complicate other modeling features (e.g., model-fit tools, variable working speed).

The {LNIRT} package provides two options to identify the model. For both options, the variance of the latent scales are identified by restricting the product of discriminations and of the time discriminations to one, $\prod_k a_k = 1$ and $\prod_k \phi_k = 1$, respectively. For option one (referred to as (R-code) {ident}=1), the mean of the scales are identified by restricting the sum of the difficulty and of the time-intensity parameters to zero, $\sum_k b_k = 0$ and $\sum_k \lambda_k = 0$, respectively. For option two (referred to as (R-code) {ident}=2), the mean of the scales are identified by fixing the mean of the ability parameter to zero, $\mu_{\theta}=0$, and of the speed parameter to zero, $\mu_{\zeta}=0$.

Model parameters are estimated through Gibbs sampling from their joint posterior distribution. The procedure involves the division of all unknown parameters into blocks, with iterative sampling of the conditional posterior distributions of the parameters in each block given the preceding draws for the parameters in all other blocks \citep{fox2010}. In various simulation studies and real data analysis, the Gibbs samplers in the {LNIRT} package showed good convergence properties and generally returned efficient MCMC samples \citep[e.g.][]{Fox2016joint,Fox2017person,kleinentink2008}.

\subsection{Logistic Versus Probit Model Item Parameter Estimates}

The {LNIRT} package uses the normal-ogive IRT model (Probit model) to model the probability of a correct/positive response. The item parameter estimates under the normal ogive (Probit) model can be transformed to a Logistic scale and vice versa. Therefore, the logistic scale factor is used (1.7) to transform parameters on a Logistic scale to those on a Probit scale. Let $\mu_L$, $\sigma_L$ and $\mu_P$, $\sigma_P$ denote the mean and variance of the latent scale under the Logistic and Probit model, respectively. Then, the item parameters $a_k$ and $b_k$ are invariant under both scales, when applying the logistic scale factor
\begin{eqnarray}
\underbrace{a_k\frac{\left(\theta_i - \mu_P\right)}{\sigma_P} - b_k}_{Probit\,Model} & = & \underbrace{1.7\left(a_k\frac{\left(\theta_i - \mu_L\right)}{\sigma_L} - b_k\right)}_{Logistic\,Model}.
\end{eqnarray}
Assume that data were generated under the Logistic IRT model, and item parameter estimates were obtained under the Probit model. The Probit model estimates can be transformed to the Logistic  model estimates;
\begin{eqnarray}
\frac{\hat{a}_k}{\sigma_P} & = & \frac{1.7a_k}{\sigma_L} \nonumber
\end{eqnarray}
and, it follows that
\begin{eqnarray}
\frac{\left(\frac{\hat{a}_k}{\sigma_P}\right)}{1.7} & = & \frac{a_k}{\sigma_L}. \nonumber
\end{eqnarray}
In the same way, the difficulty parameters under the Probit model can be transformed. When assuming that $\mu_P$ and $\mu_L$ are zero, the transformation is
\begin{eqnarray}
\hat{b}_k & = & b_k\nonumber \\
1.7\hat{b}_k & = & b^*_k, \nonumber
\end{eqnarray}
since $b_k^*=1.7b_k$ is the difficulty parameter under the Logistic model.

\section{Model Fit} \label{sec:modelfit}

\subsection{Person Fit}
Tools for evaluating the fit of the model are introduced following the procedure of  \citet{Marianti14} and \citet{Fox2017person}. The general idea is (1) to define a person-fit statistic for an RA and RT pattern, (2) to define classification variables, as a function of the statistics, to represent a non-aberrant and an aberrant state, and (3) to compute the posterior probability of the aberrant state. A person‐fit test for RA and RT patterns is discussed. For each test, a dichotomous classification variable is introduced, which states whether a pattern is considered extreme. Then, the person‐fit test for the joint model is constructed from the dichotomous classification variables for RA and RT patterns.

\subsubsection{Person-fit Statistic for RA Pattern} \label{sec:personfitRA}
\cite{drasgow1985} proposed a standardized version of Levine-Rubin statistic, which is shown to have statistical power to detect aberrant response patterns \citep{karabatsos2003}. The log-likelihood is used to evaluate the fit of an RA pattern. Following a two-parameter IRT model, the person-fit statistic, denoted as $l_\mathbf{0}$, is given by,
\begin{eqnarray}
l^{y}(\mathbf{\theta}_{i},\mathbf{a},\mathbf{b};\mathbf{y}_i) &=& -{\ln}p(\mathbf{y}_i \mid \mathbf{\theta}_{i},\mathbf{a},\mathbf{b})
\nonumber \\ &=& -\sum_{k=1}^{K} \left[\mathbf{y}_{ik}{\ln}p(\mathbf{y}_{ik}) + (1-\mathbf{y}_{ik}){\ln}(1-p(\mathbf{y}_{ik}))\right]
\end{eqnarray}
where $p(\mathbf{y}_{ik}){\equiv}p(\mathbf{y}_{ik}=1 \mid \theta_i, a_k, b_k)$. The person-fit statistic can be standardized and this standardized version, denoted as $l_s^{y}$,  is approximately standard normally distributed. Under the 3PL model, a dichotomous classification variable $S_{ik}$ is introduced that classifies a correct response to be either a correct random guess with probability $c_i$ $(S_{ik}=0)$ or a correct response according to the two-parameter IRT model $(S_{ik}=1)$. Then, $l_s^y$ is defined conditionally on $S_{ik}=1$ to evaluate the extremeness of non-guessed responses in the RA pattern. The guessed responses are ignored in the evaluation of the extremeness of an RA pattern.

To quantify the extremeness of an RA pattern, the posterior probability is computed that the person-fit statistic is greater than a threshold $C$, which can be defined according to its standard normal distribution. Note that the logarithm-of-response-probabilities are negative, thus increasing values of the negative log-likelihood correspond to misfit. The statistic is integrated over the prior distributions of all parameters and can therefore be interpreted as a prior predictive test. The (marginal) posterior probability of the statistic being greater than the threshold $C$ is expressed as;
\begin{eqnarray}
P\left(l_s^y(\mathbf{y}_i) > C\right) &=& \int ...
\int P\left(l_s^y(\mathbf{\theta}_i,\mathbf{a},\mathbf{b};\mathbf{y}_i)>C\right)
p(\mathbf{a},\mathbf{b})d\mathbf{\theta}_id\mathbf{a}d\mathbf{b}
\nonumber \\ &=& \int ...
\int \varphi\left(l_s^y(\mathbf{\theta}_i, \mathbf{a},\mathbf{b};\mathbf{y}_i)>C\right)
p(\mathbf{a},\mathbf{b})d\mathbf{\theta}_id\mathbf{a}d\mathbf{b},
\end{eqnarray}
where $\varphi$ denotes the normal density function. This approach corresponds to the prior predictive approach to hypothesis testing advocated by \citet{box1980}, since the posterior probability of an extreme RA pattern is computed by integrating over the prior distributions of the model parameters.

By introducing a classification variable, the posterior probability of an aberrant RA pattern for a given significance level can be computed. Let $F_i^\mathbf{y}$ denote a random variable that takes the value one when an observed pattern $\mathbf{y}_i$ is marked as extreme and zero otherwise,
\begin{eqnarray}
F_i^y & = & \left\{
\begin{array}{l}
1 \quad \textrm{if} \quad P\left(l_s^{y} \left(\mathbf{\theta}_{i},\mathbf{a},\mathbf{b};\mathbf{y}_i\right) > C \right), \\
0 \quad \textrm{if} \quad P\left(l_s^{y} \left(\mathbf{\theta}_{i},\mathbf{a},\mathbf{b};\mathbf{y}_i\right) \le C \right).
\end{array}
\right. \label{FlagaberrantRA}
\end{eqnarray}
The posterior probability of an extreme pattern $(F_i^\mathbf{y}$ equals one) is computed by integrating over the model parameters. Then, for a fixed critical level $C$, the posterior probability represents how likely it is that the RA pattern is extreme under the model.

\subsubsection{Person-fit Statistic for RT Pattern} \label{sec:personfitRT}
Analogously, the log-likelihood of the RT pattern of a test taker $i$ is used to define a person-fit statistic to quantify the extremeness  of the pattern. The person-fit statistic, based on the log-likelihood of an RT pattern, is represented by
\begin{eqnarray}
l_i^t\left(\zeta_i,\bm{\lambda},\bm{\phi},\bm{\sigma}^2;\mathbf{rt}_i\right) &=& \sum_{k=1}^{K} \left(\mathbf{rt}_{ik} - \left(\lambda_k - \phi_k\zeta_i \right)\right)^2/\sigma^2_k.\label{loglikeRT}
\end{eqnarray}
The sum of standardized errors is an increasing function of the negative log-likelihood, and an unusually  large person-fit value corresponds to a misfit. The $l_i^t\left(\zeta_i,\bm{\lambda},\bm{\phi},\bm{\sigma}^2;\mathbf{rt}_i\right)$ given the model parameters is chi-squared distributed with $K$ degrees of freedom. For a threshold $C$, representing the boundary of a critical region, the posterior probability of an extreme RT pattern is expressed as
\begin{eqnarray}
P\left(l_i^t\left(\zeta_i,\bm{\lambda},\bm{\phi},\bm{\sigma^2};\mathbf{rt}_i\right) > C \right) &=& P\left(\chi_{K}^2 > C \right) = p_{l^t}. \label{problt}
\end{eqnarray}

A classification variable can be defined to quantify the posterior probability of an extreme RT pattern given a threshold value $C$. Let $F_{i}^{t}$ denote the random variable which equals one when the RT pattern is flagged as extreme and zero otherwise;
\begin{eqnarray}\label{FlagaberrantRT}
F_i^t &=& \left\{
\begin{array}{l}
1 \quad \textrm{if} \quad P\left(l_i^t(\zeta_i,\bm{\lambda},\bm{\phi},\bm{\sigma}^2;\mathbf{rt}_i) > C\right), \\
0 \quad \textrm{if} \quad P\left(l_i^t(\zeta_i,\bm{\lambda},\bm{\phi},\bm{\sigma}^2;\mathbf{rt}_i) \leq C\right).
\end{array}
\right.
\end{eqnarray}
The posterior probability of an extreme RT pattern is computed for each pattern with MCMC.

\subsubsection{Person-fit Statistic for RA and RT Pattern} \label{sec:personfitRART}
An observed RT pattern $\mathbf{rt}_i$ is reported as extreme when the $F_{i}^{\mathbf{t}}$ equals one with a least .95 posterior probability. To identify a joint pattern of RA and RT to be extreme, another classification variable is defined. Let $F^{t,y}_i$ equal one, when both $F_{i}^{\mathbf{y}}$ and $F_{i}^{\mathbf{t}}$ are equal to one, and equal zero otherwise. This joint classification variable represents the situation that both patterns of a test taker are extreme or not. The classification variable for the joint pattern is defined as
\begin{eqnarray}
F^{t,y}_i &=& \left\{
\begin{array}{l}
1 \quad \textrm{if} \quad P\left(l_i^t\left(\zeta_i,\bm{\lambda},\bm{\phi},\bm{\sigma}^2;\mathbf{rt}_i\right) > C, l_s^y\left(\mathbf{\theta}_i,\mathbf{a},\mathbf{b};\mathbf{y}_i\right) > C \right), \\
0 \quad \textrm{if} \quad 1-P\left(l_i^t\left(\zeta_i,\bm{\lambda},\bm{\phi},\bm{\sigma}^2;\mathbf{rt}_i\right) > C, l_s^y\left(\mathbf{\theta}_i,\mathbf{a},\mathbf{b};\mathbf{y}_i\right)>C \right).
\end{array}
\right.\label{FlagaberrantRART}
\end{eqnarray}

The posterior probabilities of the classification variables in Equation (\ref{FlagaberrantRA}), (\ref{FlagaberrantRT}), and (\ref{FlagaberrantRART}) are Bayesian significance probabilities. They represent the posterior probability of an extreme person-fit statistic given the data. They are computed using MCMC taking into account dependencies in the joint model parameters.

\subsection{Item Fit} \label{sec:itemfit}
Without reproducing the equations for item-fit statistics, the person-fit tests for RA and RT patterns can be modified to examine the fit of RA and RT item patterns. Therefore, the log-likelihood of RA and of RTs of an item is considered to define an item-fit statistic for RA and RTs, respectively. The estimated posterior probability of each item-fit statistic represents the probability that the statistic is extreme under the model, which means that the pattern of responses to the item is extreme.

\subsection{Residual Analysis} \label{sec:resid}
Bayesian residual computation has been considered by \cite{albert1993bayesian}, \cite{johnson2006ordinal}, and \citet[][Chapter 5]{fox2010} to evaluate the fit of an IRT model. This approach is extended to the joint model, and latent residuals $e_{ik}$ are defined, which represent the difference between a latent continuous RA and the mean component. The latent continuous RA is defined through a data augmentation method to facilitate a Gibbs sampling algorithm \citep[][Chapter 3 and 4]{fox2010}. Conditional expectation of a latent residual  is derived by integrating out the augmented latent variable to obtain a Rao-Blackwell estimator for the latent residuals. The estimated latent residuals are used to quantify the total percentage of outliers per item and per test taker.

For the RA, the conditional expected latent residual is given by
\begin{eqnarray}
E\left(e_{ik} \mid a_k, b_k, \theta_i \right) & = & \left\{
  \begin{array}{ll}
\frac{-\varphi\left(b_k - a_k\theta_i \right)}{\Phi\left(b_k - a_k\theta_i\right)} & y_{ik} = 0 \\
\frac{\varphi\left(b_k - a_k\theta_i \right)}{\Phi\left(b_k - a_k\theta_i\right)}  & y_{ik} = 1 \\
  \end{array}
\right.
 \end{eqnarray}
where $\varphi$ and $\Phi$ are the normal density function and the cumulative distribution function, respectively. Subsequently, the posterior probability that a latent residual is greater than a threshold $C$ is given by
\begin{eqnarray}
P\left(\left|e_{ik} \right| > C \mid a_k, b_k, \theta_i \right) & = & \left\{
\begin{array}{ll}
\frac{\Phi\left(C \right)}{1-\Phi\left( a_k\theta_i- b_k\right)} & y_{ik} = 0 \\
\frac{\Phi\left(-C \right)}{\Phi\left(a_k\theta_i-b_k\right)} & y_{ik} = 1 \\
  \end{array}
\right.\label{extremeRA}
 \end{eqnarray}

The residuals for the RTs, referred to as $\epsilon_{ik}$, can be estimated directly as the difference between the $RT_{ik}$ and the mean, $\epsilon_{ik} = (rt_{ik} - (\lambda_k - \phi_k\zeta_i))$. The extremeness of an RT residual is expressed as the posterior probability that the residual is greater than a threshold $C$. This posterior probability can be expressed as
\begin{eqnarray}\label{extremeRT}
P\left(\left| \epsilon_{ik}\right| > C \mid \zeta_i, \lambda_k,\phi_k,rt_{ik} \right) & = & \Phi\left(-C-\frac{\epsilon_{ik}}{\sigma_k} \right) + 1-\Phi\left(C-\frac{\epsilon_{ik}}{\sigma_k} \right).
\end{eqnarray}

\subsubsection{Distribution RT Residuals}
The distribution of the RT residuals is evaluated using the Kolmogorov-Smirnov (KS) test. The empirical distribution of the residuals is compared to the assumed normal distribution. The KS test is a goodness-of-fit test and the posterior probability is computed that RT residuals of an item are non-normally distributed. The empirical distribution is given by
\begin{eqnarray}
F_N(\epsilon) & = & \frac{1}{N}\sum_{i=1}^{N} I\left(\epsilon_{ik}< \epsilon \right)(\epsilon_{ik}). \nonumber
\end{eqnarray}
The implemented KS test represents the difference between the cumulative empirical distribution and the normal cumulative distribution function:
\begin{eqnarray}
D_N & = & \sup_{\epsilon} \left| F_N(\epsilon) - \Phi(\epsilon)\right|.
\end{eqnarray}
The distribution of $D_N$ is the Kolmogorov distribution, which is used to compute the probability that $D_N$ is greater than a threshold $C$:
\begin{eqnarray}
p_{KS} & = & P\left(D_N > C \mid \mathbf{rt}_k,\lambda_k,\phi_k,\bm{\zeta} \right). \label{Pkstest}
\end{eqnarray}
The marginal posterior probability is computed using MCMC, and the estimated significance probability represents the probability that the RT residuals of item $k$ are non-normally distributed.

\subsection{Differential Working Speed}\label{sec:diffworkingspeed}
Differential working speed is defined as a change in working speed of a test taker during the test. This relaxes a basic assumption of the joint model to work with a constant speed throughout the test. The change in working speed is modeled using a latent growth model. A random intercept, a linear trend component, and a quadratic time component are considered to model the speed trajectory of each test taker. The random intercept, $\zeta_{i0}$, represents the initial value of working speed. The linear trend component, $\zeta_{i1}$, is used to model a linear change in speed. Test takers can start slowly (fast) to increase (decrease) their speed later on. The quadratic time component, $\zeta_{i2}$, is used to decelerate or accelerate the linear trend. For instance, a positive linear trend in working speed can be decelerated by a negative quadratic component.

The log-normal differential speed model with a random trend and quadratic time variable is represented by
\begin{eqnarray}\label{differentialws}
RT_{ik} & = & \lambda_k - \phi_k\left(\zeta_{i0} + \zeta_{1i}X_{ik} + \zeta_{2i}X^2_{ik} \right) + \epsilon_{ik} \\
\zeta_{i0} &\sim& N\left(0,\sigma^2_{\zeta_0} \right) \nonumber \\
\zeta_{i1} &\sim& N\left(0,\sigma^2_{\zeta_1} \right) \nonumber \\
\zeta_{i2} &\sim& N\left(0,\sigma^2_{\zeta_2} \right), \nonumber
\end{eqnarray}
where  the time variable $X_{ik}$ represents the order in which the test items are solved, where zero represents the beginning of the test. The speed process is modelled over time on an equidistant scale to reduce the MCMC computations. Speed measured at the first item represents the intercept. Let $\mathbf{X}_{(i)}=X_{(i1)},X_{(i2)},\ldots,X_{(iK)}$ represent the order in which items are solved by test taker $i$. Then, a convenient time scale is defined by $X_{ik}=(X_{(ik)}-1)/K$ -- the times are defined on a scale from 0 to 1, with 1, the upper bound, representing an infinite number of items. The scale on which working speed is measured is arbitrary. The time scale should only address the order in which the items are made and the equidistant property of the measurements.

The random effects have a population mean of zero. Thus, the average of time intensities defines the average time to complete the test. The population-average speed trajectory is constant, and shows no change in speed. Test takers can work faster (slower) than this population-average level, which corresponds to a positive (negative) initial speed level. A negative (positive) growth rate shows a decrease (increase) in speed, which can be decelerated (accelerated) by the quadratic time component. The random component variances represent the variance in growth parameters of the working speed trajectories. The covariance between random speed effects is modeled through the time-discrimination parameters. When the time-discrimination parameters are all fixed to one, the full covariance matrix of the random speed components is freely estimated.

\subsubsection{Ability and Differential Working Speed}
A multivariate normal distribution is assumed for the random component ability and speed to allow for relationships between ability and the different speed components. This multivariate model for the random person parameters, $(\theta_i,\bm{\zeta}_i)=(\theta_i,\zeta_{0i},\zeta_{1i},\zeta_{2i})$, is given by
\begin{eqnarray}
\left(
  \begin{array}{l}
    \theta_i                \\
    \bm{\zeta}_i       \\
  \end{array}
\right)
& = &
N_4\left(\left(
   \begin{array}{l}
    \mu_{\theta}    \\
    \bm{0}     \\
  \end{array}
\right),
\left(
   \begin{array}{ll}
    \sigma^2_{\theta}   & \bm{\Sigma}_{\theta,\zeta} \\
    \bm{\Sigma}_{\zeta,\theta} & \bm{\Sigma}_{\zeta}    \\
  \end{array}
\right)\right).
\end{eqnarray}
The relationship between ability and the speed components is defined by the covariance components $\bm{\Sigma}_{\theta,\zeta}$. The growth components defining the speed trajectory influence ability. When test takers do not vary speed, ability is only influenced by the random intercept speed. When test takers vary their speed, the trend and quadratic time component will also influence ability, which is a specific feature of this generalization of the constant speed model.

Whether a change in speed improves the accuracy of the responses depends on the application. The differential speed joint model can be used to examine different test strategies across test takers. For instance, it is possible to estimate the speed trajectories of test takers with different levels of ability. The speed trajectories of high-ability students can differ from the low-ability students. The speed trajectories of test takers may also differ across tests. The model can be used to explore effects of time limits on test takers' changes in working speed. Benefits of exploring heterogenous speed trajectories in relation to ability will depend on the application.

\section{Software}
The main function of the package {LNIRT} is the \textit{LNIRT} function to fit the joint model for RA and RTs. It checks the input, arranges the input for the MCMC algorithm, and constructs the data output. An object of class \textbf{LNIRT} is generated, and a summary function can be used to get a summarized view of the estimation results. In the most simple case, the user passes the data to the \textit{LNIRT} function and uses the package's summary function to get an overview of the results. The complete functionality of the package can be accessed by making further input specifications.

\subsection{Input}
The following arguments are mandatory for the \textit{LNIRT} function:
\begin{itemize}
\item RT: A matrix RT containing the \textit{log-normally} transformed response times in wide format, (N) persons in rows and (K) items in columns.
\item Y: A matrix Y containing the (binary) RA data in wide format, (N) persons in rows and (K) items in columns. The binary RA data is coded as zero (incorrect) and one (correct) (missing values as NA).
\item data (optional if RT and Y are given): A list or a \textit{simLNIRT} object (output object of the function \textit{simLNIRT}) containing the RT and RA matrices and optionally predictors for the item and person parameters. If a \textit{simLNIRT} object is provided, in the summary the simulated item and time parameters are shown alongside of the estimates. If the required variables cannot be found in the list, or if no data object is given, then the variables are taken from the environment from which \textit{LNIRT} is called.
\item XG: The integer number of MCMC iterations (the default is 1,000), this includes the burn-in period.
\end{itemize}
The remaining arguments are optional for the \textit{LNIRT} function:
\begin{itemize}
\item burnin: The percentage of the total number of MCMC iterations (XG) which will serve as the burn-in period of the chains. The default is 10\%.
\item ident: Identification rule, (ident=1) restrict sum of difficulties and sum of time intensities to zero and (ident=2) restrict mean person parameters to zero. The default is (ident=2). The product of (time) discriminations is restricted to one.
\item guess: A logical variable with the default FALSE, where TRUE (FALSE) represents (not) a guessing parameter in the IRT model.
\item par1: A logical variable with the default FALSE, where TRUE represents the bracket notation for the mean term of the IRT component as in Equation (\ref{eq:responsemodel1}), and FALSE the non-bracket notation as in Equation (\ref{eq:responsemodel}). In general, the MCMC performance is better for the non-bracket parameterization.
\item XGresid: the number of MCMC iterations (default is 1,000) to be done before starting the residual computation.
\item residual: A logical variable with the default FALSE. When TRUE, a complete residual analysis is done together with the estimation of the joint model parameters, as discussed in Section \ref{sec:modelfit}. The residual computations are started after XGresid MCMC draws. Therefore, XG should be greater than XGresid, and a sufficient number of MCMC iterations should be made after XGresid MCMC iterations to obtain accurate residual estimates. Preferably, at least 5,000 MCMC iterations are made, when doing a residual analysis.
\item td: A logical variable with the default TRUE. When TRUE, the time-discrimination parameter is estimated. When FALSE, the time discrimination is restricted to one.
\item WL: A logical variable with the default FALSE. When TRUE, the time-discrimination parameter represents the inverse of the measurement error variance parameter according to the parameterization of \cite{Linden2007}.
\item alpha: An optional vector of length K of pre-defined item discrimination parameters.
\item beta: An optional vector of length K of pre-defined item difficulty parameters.
\item phi: An optional vector of length K of pre-defined time-discrimination parameters.
\item lambda: An optional vector of length K of pre-defined time-intensity parameters.
\item XPA: An optional matrix of predictor variables for the ability parameters, where the columns represent the predictor variables. Categorical predictor variables need to be dummy coded.
\item XPT: An optional matrix of predictor variables for the speed parameters, where the columns represent the predictor variables. Categorical predictor variables need to be dummy coded.
\item XIA: An optional matrix of predictor variables for the item difficulty parameters, where the columns represent the predictor variables. Categorical predictor variables need to be dummy coded.
\item XIT: An optional matrix of predictor variables for the time-intensity parameters, where the columns represent the predictor variables. Categorical predictor variables need to be dummy coded.
\item MBDY: An optional missing-by-design indicator matrix -- of the same size as Y -- for missing values (coded NA)  in the Y matrix due to the test design (0 is missing by design, 1 is not missing by design). Multiple imputations are simulated for missing values (missing at random) in the Y matrix which are not assigned in MBDY as missing-by-design.
\item MBDT: An optional missing-by-design indicator matrix -- of the same size as RT -- for missing values (coded NA) in the RT matrix due to the test design (0 is missing by design, 1 is not missing by design). Multiple imputations are simulated for missing values (missing at random) in the RT matrix which are not assigned in MBDT as missing-by-design.
\end{itemize}

\subsection{Output}
The \textit{LNIRT} function creates an object of class \textbf{LNIRT}, which stores the MCMC output, posterior draws and posterior mean estimates. The output of the function \textit{LNIRT} is described in an itemized way, without including a residual computation.
\begin{itemize}
\item data: If available a data object from the function \textit{simLNIRT}.
\item burnin: Percentage from XG representing the burn-in MCMC iterations.
\item ident: Same as the input variable ident.
\item guess: Same as the input variable guess.
\item MAB: The MCMC sampled values for the item parameters (discrimination, difficulty, time discrimination, time intensity), object is an array of dimension XG (number of MCMC iterations) by K (number of items) by 4 (number of item parameters).
\item MCMC.Samples: This object contains the sampled MCMC values from \textit{LNIRT}, where the following objects are stored:
    \begin{itemize}
    \item Cov.Person.Ability.Speed : Samples of covariance of ability and speed.
    \item CovMat.Item : Array of dimension XG (MCMC iterations) by K (items) by 2 (time-discrimination and time-intensity parameters), it contains the sampled variance of time discrimination and time intensity as well as their sampled covariance.
    \item Item.Dificulty: Samples of difficulty parameters (XG by K).
    \item Item.Discrimination: Samples of discrimination parameters (XG by K).
    \item Item.Guessing: Samples of guessing parameters (XG by K).
    \item Mu.Item.Difficulty: Sampled values of mean item difficulty parameter.
    \item Mu.Item.Discrimination: Sampled values of mean item discrimination parameter.
    \item Mu.Person.Ability: Sampled values of mean ability parameter.
    \item Mu.Person.Speed: Sampled values of mean speed parameter.
    \item Mu.Time.Discrimination: Sampled values of mean time-discrimination parameter.
    \item Mu.Time.Intensity: Sampled values of mean time-intensity parameter.
    \item Person.Ability: Sampled values of person's ability parameter (XG by N).
    \item Person.Speed: Sampled values of person's speed parameter (XG by N).
    \item Sigma2: Sampled values of measurement error variance parameters (XG by K).
    \item Time.Discrimination: Sampled time-discrimination parameters (XG by K).
    \item Time.Intensity: Sampled time-intensity parameters (XG by K).
    \item Var.Person.Ability: Sampled variance ability parameters.
    \item Var.Person.Speed: Sampled variance speed parameters.
    \end{itemize}
\item Mguess: Sampled values for the guessing parameter (under the default Beta prior B(20,80)).
\item MmuI: Sampled values of the mean discrimination, difficulty, time discrimination, and mean time intensity (in that order).
\item MmuP: Sampled values of the mean ability and mean speed parameters.
\item MSI: Sampled values of the covariance matrix of item parameters. Array of dimension XG by 4 by 4 (discrimination, difficulty, time discrimination, time intensity), see also CovMat.Item.
\item Msigma2: Sampled values of measurement error variance parameters.
\item MSP: Sampled values of the covariance matrix of person parameters (ability and speed), in an array of dimension XG by 2 by 2.
\item Mtheta: Posterior mean estimates of ability and speed of dimension N by 2.
\item MTSD: Posterior standard deviation of ability and speed of dimension N by 2.
\item par1: Same as the input variable par1.
\item Post.Means: Posterior mean estimates of the following parameters:
\begin{itemize}
    \item Cov.Person.Ability.Speed: Covariance ability and speed.
    \item CovMat.Item: Covariance matrix item parameters.
    \item Item.Difficulty: Item difficulty estimates.
    \item Item.Discrimination: Item discrimination estimates.
    \item Mu.Item.Difficulty: Mean item difficulty estimate.
    \item Mu.Item.Discrimination: Mean item discrimination estimate.
    \item Mu.Person.Ability: Mean ability estimate.
    \item Mu.Person.Speed: Mean speed parameter.
    \item Mu.Time.Discrimination: Mean time discrimination.
    \item Mu.Time.Intensity : Mean time intensity.
    \item Person.Ability: Ability parameters.
    \item Person.Speed: Speed parameters.
    \item Sigma2: Measurement error variance.
    \item Time.Discrimination: Time-discrimination parameters.
    \item Time.Intensity: Time-intensity parameters.
    \item Var.Person.Ability: Variance ability parameter.
    \item Var.Person.Speed: Variance speed parameter.
\end{itemize}
\item RT: Logarithm of the RT data.
\item td: Same as the input variable td.
\item WL: Same as the input variable WL.
\item XIA: Same as the input variable XIA.
\item XIT: Same as the input variable XIT.
\item XPA: Same as the input variable XPA.
\item XPT: Same as the input variable XPT.
\item Y: Same as the input variable Y.
\end{itemize}
When the residual computation is included (residual=TRUE), then the \textbf{LNIRT} object includes the following additional output variables:
\begin{itemize}
\item EAPCP1: Posterior probability that the RT pattern is flagged as aberrant according to Equation (\ref{FlagaberrantRT}), using the posterior probability that the person-fit statistic $l^t$ is extreme as defined in Equation (\ref{problt}) with a significance level of .05.
\item EAPCP2: Posterior probability that the RA pattern is flagged as aberrant according to Equation (\ref{FlagaberrantRA}), using the posterior probability that the person-fit statistic, $l^y_s$, is significant (with a significance level of .05).
\item EAPCP3: Posterior probability that both patterns (RA and RT) are flagged as aberrant according to Equation (\ref{FlagaberrantRART}).
\item EAPKS: Posterior probability of an extreme KS-test result according to Equation (\ref{Pkstest}), which indicates that the RT residuals are not normally distributed.
\item EAPKSA: A significance probability of an extreme  KS-test that the latent residuals of RA items are not normally distributed. This significance test has no power.
\item EAPresid: Posterior probability of an extreme (standardized) residual (RT data), which is greater than plus or minus two in absolute value.
\item EAPresidA: Posterior probability of an extreme (standardized) latent residual (RA data),  which is greater than plus or minus two in absolute value.
\item IFl: The (negative) log-likelihood statistic for item fit under the IRT model (high values represent misfit).
\item IFlp: The posterior significance probability of an extreme item fit under the IRT model.
\item lZI: Item-fit statistic representing the posterior probability that the squared sum of item residuals are extreme under the model.
\item EAPl0: The log-likelihood contribution of each RA observation, where a low value represents a misfit.
\item PFl: The standardized (negative) log-likelihood contribution of each RA pattern, where a high value represents a misfit.
\item PFlp: Posterior (significance) probability of observing a more extreme person-fit statistic for the RA pattern than the observed one.
\item lZPA: Posterior significance probability of an extreme person-fit test for RA pattern based on latent residuals. This significance test has no power (under construction).
\item lZPT: The (unstandardized) estimated person-fit statistic according to Equation (\ref{loglikeRT}).
\item lZP: Posterior (significance) probability of observing a more extreme person-fit statistic for the RT pattern than the observed one, according to Equation (\ref{problt}).
\end{itemize}

\subsection{MCMC Output Analysis}
The posterior draws in the output of {LNIRT} are obtained through a Gibbs sampler. MCMC convergence tests are necessary to make sure that the MCMC chains converged before making statistical inferences. The {coda} package \citep{plummer2006coda} can be used to do an MCMC convergence analysis. The {LNIRT} function returns a single MCMC chain for each model parameter, which can be directly translated to an \textbf{MCMC} object with the function \textit{as.mcmc}. The single-chain convergence diagnostics, for instance Geweke, and Heidelberg-Welch, can be used to examine if the chains did not converge. The multiple-chain convergence diagnostics, for instance Gelman and Rubin's convergence statistic, requires multiple calls to {LNIRT} to create multiple MCMC chains with different starting values -- {LNIRT} uses random starting values when parameter values are not pre-specified.

With the summary function of the package {LNIRT}, posterior means and standard deviations are computed and reported. Although the {LNIRT} Gibbs samplers produce efficient MCMC samples with low autocorrelation, it is possible that for a specific dataset some of the chains have a medium to high autocorrelation. Then, the reported standard deviation estimates may underestimate the standard deviation, since autocorrelation in the chains are ignored. The {coda} and the {mcmcse} package can be used to compute the autocorrelation and the Monte Carlo standard error. Note that the autocorrelation has no effect on the posterior mean estimate. The effective sample size -- the sample size of independently distributed values with the same variance as the autocorrelated MCMC sample -- can also be computed to examine if the run was long enough to make accurate and reliable inferences. A reasonable rule of thumb is to have an effective sample size of 400. Then, the Monte Carlo standard error is less than 5\% of the overall uncertainty of the posterior mean.

The default burn-in period is 10\% of the total number of MCMC iterations. This burn-in period is used in the computation of the posterior estimates, which are also reported with the summary function. Extensive simulation studies showed that the burn-in period is usually below 100 MCMC iterations, but this could be higher for a specific data set. Furthermore, the MCMC properties are better for the RA model in Equation (\ref{eq:responsemodel}) than for the RA model with the additional brackets given in Equation (\ref{eq:responsemodel1}).

\section{Applications} \label{sec:application}

\subsection{The Credential (Form1) Data}
The credential data set of \cite{cizek2016handbook} is analyzed to illustrate the functionality of the package {LNIRT}. The credential data set concerns 1,636 test takers who applied for licensure. Form 1 of the test was administered, which consisted of 170 licensure exam items, and 30 pretest items. A total of 10 background variables of the test takers was stored -- this includes the country where the candidate received his/her educational training, the state in which the test taker applied for licensure, and the center where the candidate took the exam. The RA and RT data were also stored. The collected data followed from a year of testing using a computer-based program that tests continuously.

Each candidate completed one of the three pretest forms, each consisting of 10 items. In Table \ref{tab:tab2}, the test design is given, which shows the incomplete test design of in total 200 items for three groups.

\begin{table}[h!]
\centering
\begin{tabular}{ccccc}
\hline
 & \multicolumn{4}{c}{Item Set} \\ \cline{2-5}
  &   & \multicolumn{3}{c}{Pretest} \\ \cline{3-5}
Group & Test (1-170) & K1 (171-180) & K2 (181-190) & K3 (191-200) \\ \hline
G1 & X & X & - & - \\
G2 & X & - & X & - \\
G3 & X & - & - & X \\
\hline
\end{tabular}
\caption{\label{tab:tab2} Credential data (Form 1): The test design.}
\end{table}

The RA and RT data were extracted from the data to be used in the {LNIRT()} function. The RA data consisted of 1636 test takers (in rows) and 170 exam items (in columns).
\begin{verbatim}
R> Y<- as.matrix(data[c(which(colnames(data)=="iraw.1")
+	:which(colnames(Cdata)=="iraw.170"))])
R> head(Y[,1:5],3)
\end{verbatim}
\begin{verbatim}
     iraw.1 iraw.2 iraw.3 iraw.4 iraw.5
[1,]      1      1      1      0      1
[2,]      1      0      1      0      0
[3,]      0      0      0      0      1
\end{verbatim}

The RT data also consisted of 1636 test-takers and 170 exam items. The RTs were transformed to a logarithmic scale. A total of 105 RTs were equal to zero. Possibly the item was skipped and a default incorrect response was recorded, since the corresponding RA were all incorrect. The zero RTs were converted into NAs, since it was unknown why a zero RT was recorded. The minimum RT was 2 seconds, and the 100 highest RTs ranged from 6 to 12 minutes.
\begin{verbatim}
R> RT<-as.matrix(data[c(which(colnames(data)=="idur.1"):
+	which(colnames(data)=="idur.170"))])
R> RT[RT==0]<-NA
R> RT<-log(RT)
R> head(RT[,1:5],3)
\end{verbatim}
\begin{verbatim}
       idur.1   idur.2   idur.3   idur.4   idur.5
[1,] 4.094345 3.555348 3.555348 3.465736 3.295837
[2,] 4.007333 4.060443 4.343805 3.850148 4.110874
[3,] 4.234107 3.761200 4.007333 3.178054 4.127134
\end{verbatim}

In the first analysis, the (default) joint model was fitted to the data to explore the item parameter estimates and the item and person covariance matrix. The model was identified by restricting the population means of ability and speed to zero and by restricting the product of time discriminations and discriminations to one. A total of 5,000 MCMC iterations were computed, and the burn-in period was 500 (i.e. 10\% of the total number of MCMC iterations).
\begin{verbatim}
R> library(LNIRT)
R> out0 <- LNIRT(RT=RT, Y=Y, XG=5000,ident=2,burnin=10)
R> summary(out0)
\end{verbatim}

\subsubsection*{MCMC Convergence}
Multiple MCMC chains were checked for convergence and run length by computing the Geweke and Heidelberger statistics, the effective sample sizes, and the MCMC standard errors. The convergence statistics did not show any problems of non-convergence for the examined chains. The lowest effective sample size was 427 for the discrimination parameter of item 155 (MCMC standard error of .007), and the highest 4,000 for the speed parameter of test taker 1 (MCMC standard error less than .000). The naive posterior standard deviation -- when ignoring autocorrelation in the chain -- was often just smaller than the times-series standard error estimate using the {coda} package. It was concluded that the burn-in period and the total length of the chains were sufficiently long to compute accurate parameter estimates.

\subsubsection*{Output Analysis}
\begin{figure}[ht!]
    \centering
    \includegraphics[width=\linewidth]{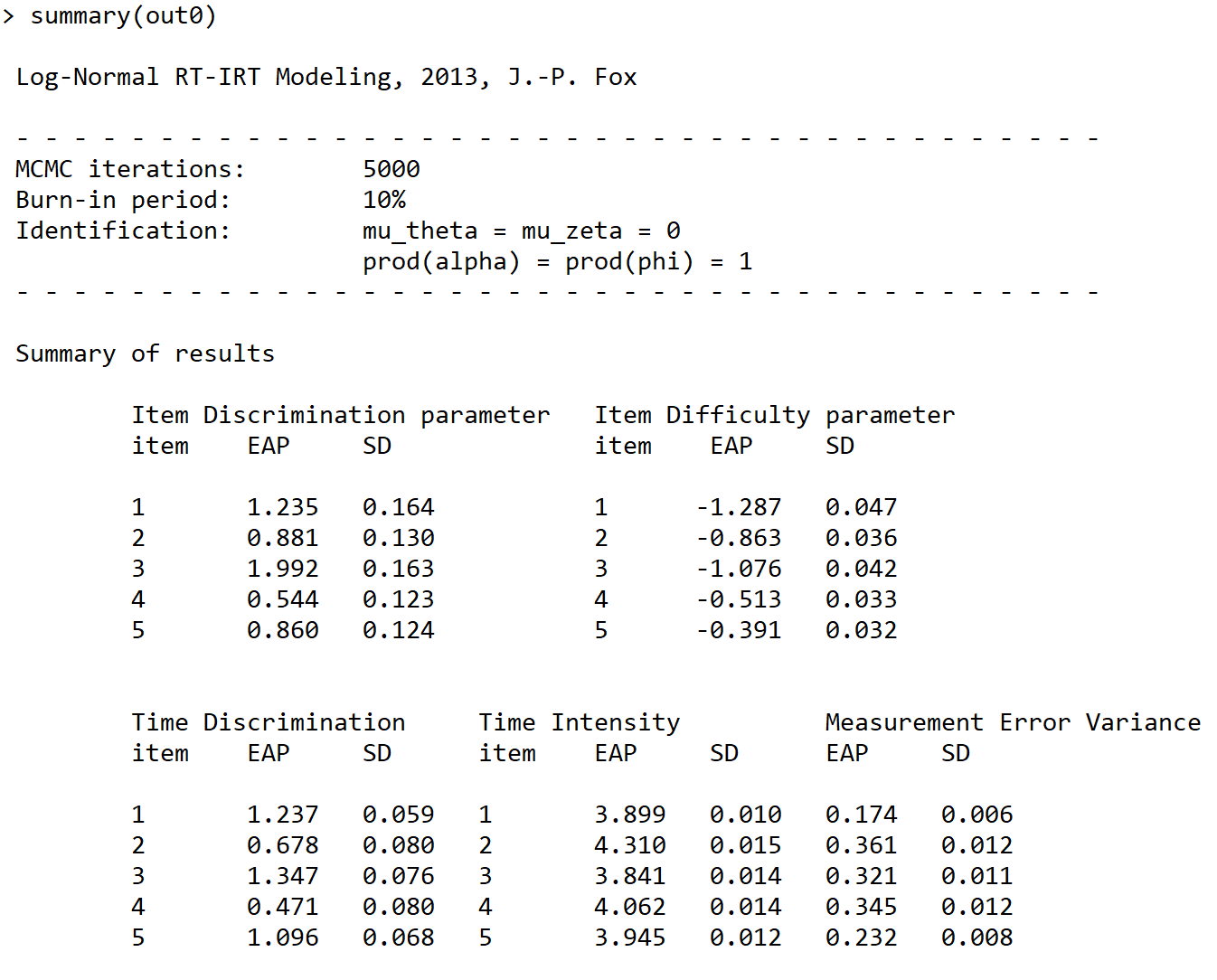}
     \caption{Screenshot of the {LNIRT} summary output: Item parameter estimates of the first five items.}
     \label{fig:screenshotitems}
  \end{figure}

The general summary function of {LNIRT} shows the item number (which corresponds to the column number of Y and of RT), and the posterior mean (labeled EAP) and standard deviation (labeled SD) estimates of the item and of the prior parameters of the items and persons. In Figure \ref{fig:screenshotitems}, the estimation results for the first five item parameters are given.

\begin{figure}[ht!]
    \centering
    \includegraphics[width=\linewidth]{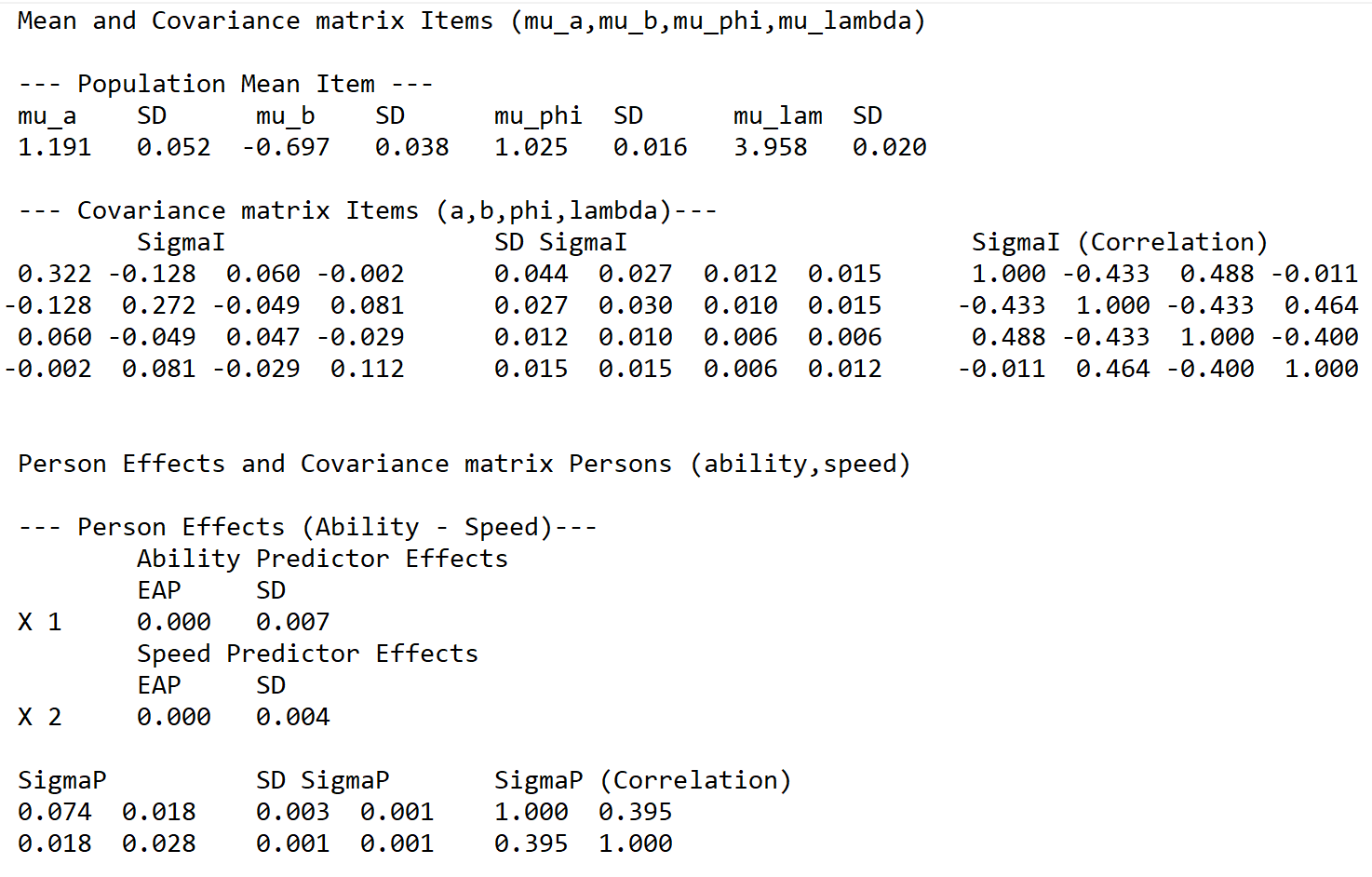}
     \caption{Screenshot of the {LNIRT} summary output: Prior parameter estimates of persons and items.}
     \label{fig:screenshotpriors}
  \end{figure}

In Figure \ref{fig:screenshotpriors}, the item and person prior parameter estimates are given. The mean of the items, $\mu_I$, is labeled in the output as mu\_a (mean discrimination), mu\_b (mean difficulty), mu\_phi (mean time discrimination) and mu\_lam (mean time intensity). The posterior mean estimate of the covariance matrix of the items, $\bm{\Sigma}_I$, is given under the label Sigma\_I. The estimated mean item discrimination is around 1.19 with a variance of around .32. Around 10 items have an estimated item discrimination below .40. The item difficulty estimates ranges from -1.84 to .75, with a mean of -.70 and a variance of .27. The estimated average time-discrimination and time-intensity is around 1.03 and 3.96, with a variance of .05 and .11, respectively. The time discrimination ranged from .47 to 1.42, and it appears that the items show better discriminating performance in speed than in ability. However, the estimated working-speed variance is very small and around .03. The variance in RTs is mostly explained by the time-intensity parameters and hardly by differences in working speed across test takers. Thus, the time discriminations are somewhat higher than the item discriminations, but the time discriminations have a minor contribution in explaining variance in RTs.

The covariance matrix of the item parameters shows that the less difficult items are more discriminating (correlation of -.43), and the less time-intensive items are also more time-discriminating (correlation of -.40). Differences in ability and speed are better measured with less difficult and less time-intensive items. The more difficult items are also more time-intensive (correlation of .46). The time-discriminating items are positively correlated with the item discriminations (correlation of .49).

The covariance matrix of the person parameters (ability and speed), $\bm{\Sigma}_P$, is given under the label Sigma\_P. The variance in ability and speed across test takers are both small.  There is a positive correlation between ability and speed of around .40, which states that more able test takers also work faster than the less able ones.

\subsubsection*{Explanatory Variables}
Dummy coded variables were created for the pretest groups, where effect coding is used such that the general intercept can be restricted to zero to identify the joint model. Differences in ability and speed were examined across pretest groups (stored in objects XA and XT for ability and speed), and the test-taker's total test time was used to explain the correlation between ability and speed (stored in object XA for ability). The
observed total test times contained information on top of the speed values, since the latter were shrunk towards the population-average test time depending on the variance of the speed parameters. The joint model parameters -- with the explanatory variables for ability and speed -- are estimated. The main estimation results are again  computed and displayed using the (R-code) {summary} function (using defaults burnin=10 and ident=2):
\begin{verbatim}
R> XFT$Pgroup[XFT$Pretest==6,1] <- -1
R> XFT$Pgroup[XFT$Pretest==6,2] <- -1
R> XFT$Pgroup[XFT$Pretest==7,1] <- 1
R> XFT$Pgroup[XFT$Pretest==8,2] <- 1
R> out1 <- LNIRT(RT=RT, Y=Y, XG=5000, XPA=XA, XPT=XT)
R> summary(out1)
\end{verbatim}

The effect of the total test time on ability was significant and around -.07. Those who finished earlier scored on average higher than those who finished the test later. When accounting for the total test-time differences in measuring ability, the correlation between ability and speed was around .1. So, the total test-time explained around 75\% of the correlation. There were no significant speed differences between the pretest groups, and the variance of speed was also small and around .03. There were estimated differences in ability between the pretest groups: group $G_1$ scored on average -.13 lower, group $G_2$ .07 higher, and  group $G_3$ .05 higher than the general average, but they were not significant.

\subsubsection*{Planned Missing by Design}
In the planned missing data design, simultaneous parameter estimation for all examinees is possible using (R-code) {LNIRT()} function. It requires defining an indicator matrix for the planned missing data. In the matrix {MBDM}, a zero is a designed missing and a one a designed observation. The planned missing data matrix can be defined separately for the RA and RT data, and arguments {MBDY} and {MBDT} represent the design matrix for the RA and RT data, respectively. For instance, it is possible to include RA data in the analysis for which RT data was only partly collected. When including the pretest items in the measurement of ability and speed, the test design contains planned missing data. Then, the indicator matrix {MBDM} for the planned missing data is defined and included in the call to {LNIRT}:
\begin{verbatim}
R> MBDM<-matrix(rep(0,1636*200),nrow=1636,ncol=200)
R> MBDM[XFT$Pretest==6,171:180]<-1
R> MBDM[XFT$Pretest==7,181:190]<-1
R> MBDM[XFT$Pretest==8,191:200]<-1
R> MBDM[,1:170]<-1
R> outmbdm <- LNIRT(RT=RTt,Y=Yt,XG=5000,alpha=alpha1,MBDY=MBDM,MBDT=MBDM)
\end{verbatim}

In the call to {LNIRT}, the arguments {RTt} and {Yt} contain the RT and RA data for all 200 items. Pre-specified item discriminations, {alpha1}, are used, since the model with free item discrimination parameters would not fit. The output is not discussed for reasons of brevity.

\subsubsection*{Model Fit}
The joint model-fit tools are illustrated by re-running the analysis with explanatory variables for the 170 items and activating the residual computation (R-code {residual=TRUE}).
\begin{verbatim}
R> out1 <- LNIRT(RT=RT,Y=Y,XG=5000,XPA=XA,XPT=XT,residual=TRUE)
R> summary(out1)
\end{verbatim}

The summary report contains an overview of the extreme residuals under the header \textit{Residual Analysis}. A total of .07\% of RTs residuals were considered extreme with at least 95\% posterior probability (Equation (\ref{extremeRT})). This concerns RTs that were small and around 2-6 seconds or much higher than the item-average RTs. For the RA residuals, around .02\% was estimated to be extreme with 95\% posterior probability (Equation (\ref{extremeRA})). This concerns incorrect responses from test takers with an above-average ability. For 62 items (36.5\%) the assumption of log-normally distributed residuals was violated for which a significant probability of the KS test was computed (Equation (\ref{Pkstest})). The variance in working speed and time intensities are small, and the estimated residual variance is around .26. Therefore, RT outliers more easily affect the fit of the log-normal distribution. The item-fit statistics (RA and RT data) did not identify a significant misfit of an item. Despite the outliers, log-likelihoods of item patterns were not significant under the joint model.

The {EAPCP1} is reported and around 18.34\%, representing the percentage of RT patterns which are considered extreme with 95\% posterior probability. The percentage of significant extreme RT patterns is around 19.5\%, when using a significance level of .05. The reported {EAPCP2} is around 1.59\% and the significant RA patterns is around 1.65\%, which shows that there are only a few RA patterns identified as extreme. Finally, around .31\% of the joint patterns (RA and RT) are extreme (object  {EAPCP3}).

The heterogeneity in person-fit statistics for RA an RT patterns is examined with a linear regression of the statistics on the number of test attempts, country, and pretest group. The country variable was (dummy) recoded (US (Cgroup1=1),Philippines (Cgroup2=1),India (Cgroup3=1),Others (intercept)). The pretest groups were already represented by two dummy coded variables.
\begin{verbatim}
 R> summary(lm(out1$PFl ~ as.factor(XFT$Attempt)+(XFT$Cgroup)+(XFT$Pgroup)))

Coefficients:
                          Estimate Std. Error t value Pr(>|t|)
(Intercept)              -0.204245   0.045153  -4.523 6.53e-06 ***
as.factor(XFT$Attempt)2  -0.004323   0.060106  -0.072  0.94267
as.factor(XFT$Attempt)3  -0.102401   0.079525  -1.288  0.19805
as.factor(XFT$Attempt)4  -0.253421   0.114568  -2.212  0.02711 *
as.factor(XFT$Attempt)4+  0.017670   0.080984   0.218  0.82731
XFT$Cgroup1               0.051825   0.046087   1.124  0.26097
XFT$Cgroup2              -0.215907   0.054834  -3.937 8.58e-05 ***
XFT$Cgroup3              -0.109458   0.053455  -2.048  0.04075 *
XFT$Pgroup1               0.011785   0.034349   0.343  0.73157
XFT$Pgroup2               0.095390   0.029290   3.257  0.00115 **
\end{verbatim}

It follows that the RA patterns of test takers with more test attempts are less extreme than those with a fewer attempts. Test takers from the Philippines and India are less likely to have an aberrant RA pattern. Furthermore, test takers from the third pretest group (variable Pgroup2) have higher person-fit statistic scores. They correspond to RA patterns that are less likely to be observed under the model than RA patterns with lower person-fit statistics. For the person-fit test for RA patterns, for a significance level of .05 the critical value is 1.645. The average-statistic scores (number of attempts, country, pretest group) are much lower than the critical value, and test takers with an aberrant RA pattern are also outliers in their groups.

The heterogeneity in person-fit statistic scores for the RT patterns is also explored through a linear regression with the same explanatory variables.
\begin{verbatim}
R> summary(lm(out1$lZPT ~ as.factor(XFT$Attempt)+(XFT$Cgroup)+(XFT$Pgroup)))

Coefficients:
                         Estimate Std. Error t value Pr(>|t|)
(Intercept)              169.9231     3.0855  55.071  < 2e-16 ***
as.factor(XFT$Attempt)2    1.1506     4.1073   0.280   0.7794
as.factor(XFT$Attempt)3   10.2410     5.4343   1.885   0.0597 .
as.factor(XFT$Attempt)4    2.6948     7.8290   0.344   0.7307
as.factor(XFT$Attempt)4+   2.3246     5.5341   0.420   0.6745
XFT$Cgroup1               -7.7563     3.1494  -2.463   0.0139 *
XFT$Cgroup2                3.0138     3.7471   0.804   0.4213
XFT$Cgroup3               15.4076     3.6528   4.218  2.6e-05 ***
XFT$Pgroup1                2.5152     2.3473   1.072   0.2841
XFT$Pgroup2                0.2681     2.0016   0.134   0.8935
\end{verbatim}

For the $l_i^t$ statistic, the critical value is 201.4 when the significance level is .05 ($l_i^t$ is chi-square distributed with 170 degrees of freedom). The intercept corresponds to test takers from pretest group 1 from citizens outside the US, India, and the Philippines, who took the test for the first time. For test takers with more attempts the person-fit statistic is on average higher and closer to the critical value. Those test takers are more likely to give a very fast or much slower response. Test takers from the US have on average much lower person-fit scores, and test takers from India have a much higher person-fit score.

\begin{figure}[ht!]
    \centering
    \includegraphics[width=\linewidth]{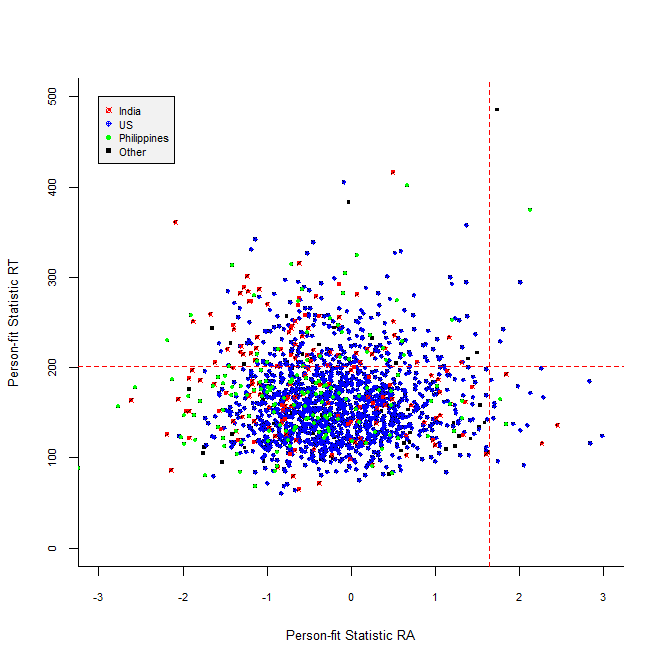}
     \caption{For different countries the person-fit statistic of RA patterns plotted against those of RT patterns.}
     \label{fig:pftestscountry}
  \end{figure}

In Figure \ref{fig:pftestscountry}, the person-fit scores for RA patterns are plotted against those of RT patterns to provide a more comprehensive overview of the aberrant and non-aberrant patterns per country. For this example, the R-code is given below
\begin{verbatim}
set1 <- which(XFT$Country=="USA")
set2 <- which(XFT$Country=="India")
set3 <- which(XFT$Country=="Philippines")
plot(out1$PFl,out1$lZPT,xlab="Person-fit Statistic RA",
    ylab="Person-fit Statistic RT",col="black",cex=.5,bty="l",xlim=c(-3,3),
    ylim=c(0,500),cex.main=.8,cex.axis=.7,cex.lab=.8,pch=15)
points(out1$PFl[set1],out1$lZPT[set1],col="blue",pch=10,cex=.5)
points(out1$PFl[set2],out1$lZPT[set2],col="red",pch=13,cex=.5)
points(out1$PFl[set3],out1$lZPT[set3],col="green",pch=16,cex=.5)
abline(h = qchisq(.95, df= 170),lty = 2,col="red")
abline(v = qnorm(.95),lty = 2,col="red")
legend(-3,500,c("India","US","Philippines","Other"), col=c("red","blue",
    "green","black"),pch = c(13,10,16,15), bg = "gray95",cex=.7)
\end{verbatim}

For both person-fit statistics, the threshold value of the significant area is marked with a dotted red line. With respect to aberrant RT patterns, a serious number of test takers are marked as aberrant, since their value is above the threshold of 201.4. Test takers from India have relatively more aberrant persons with respect to RT patterns, and the US test takers relatively less, although many of the aberrant test takers are US citizens. A few test takers are marked as aberrant with respect to their RA pattern, since their statistic value is above 1.645. Only five persons have been marked as aberrant in terms of the RA and RT patterns. None of the four quadrants seems to be dominated by one specific country.

\subsection{The Amsterdam Chess Data}

The Amsterdam Chess  data of \citet{van2005psychometric} is used to illustrate the measurement of variable speed trajectories of 259 test takers, who responded to 40 chess tasks. The chess items have three subdimensions; tactical skill (20 items), positional skill (10 items), and end-game skill (10 items). Each item represents a chess board situation, and the problem-solving task was to select the best possible move. Both RAs and RTs were recorded, where RA was coded as 1 (correct) and 0 (incorrect). In this analysis, the differential working speed model in Equation (\ref{differentialws}) is fitted to explore heterogeneity in speed trajectories and to examine the between-person relationship between ability and random speed components.

The {LNIRTQ} function is used to run the joint model with a differential working speed model with a random trend and quadratic time component. The RA and RT data were extracted from the Amsterdam Chess data to be used in the {LNIRTQ} function. The RA data consisted of 259 test takers (in rows) and 40 chess items (in columns).
\begin{verbatim}
R> data(AmsterdamChess)
R> N <- nrow(AmsterdamChess)
R> Y <- as.matrix(AmsterdamChess[c(which(colnames(AmsterdamChess)=="Y1"):
        which(colnames(AmsterdamChess)=="Y40"))])
R> K <- ncol(Y)
R> Y[Y==9] <- NA
R> RT <- as.matrix(AmsterdamChess[c(which(colnames(AmsterdamChess)=="RT1"):
        which(colnames(AmsterdamChess)=="RT40"))])
R> RT[RT==10000.000] <- NA
R> RT <- log(RT) #logarithm of RTs
\end{verbatim}

The missing RA and RT values coded as 9 and 10,000 are replaced by NAs. There are three records with all missing values (rows 147,201, and 209). In {LNIRTQ}, imputations are generated under the model for the missing data even if complete records are missing. A time scale is defined by starting at zero and taking steps of 1/K to end in (K-1)/K. The LNIRTQ model is identified by restricting the difficulty and discrimination parameters to zero and one, respectively. Furthermore, the mean of each random speed parameter (intercept, trend, quadratic) is restricted to zero, and the product of time discriminations is restricted to one. Finally, the covariance of the speed components is restricted to zero, since the covariance among the speed components is modeled by the time discriminations. The {LNIRTQ} function is ran for 10,000 MCMC iterations, with a default burnin-period of 10\%.
\begin{verbatim}
R> X <- 1:K
R> X <- (X - 1)/K
R> outchess <- LNIRTQ(Y=Y,RT=RT,X=X,XG=10000)
R> summary(outchess)
\end{verbatim}

MCMC chains can be checked in the same way for convergence and run length by computing the Geweke and Heidelberger statistics, the effective sample sizes, the MCMC standard errors. The output of {LNIRTQ} contains MCMC chains of item parameters and hyper prior parameters for items and persons. The examined chains did not have any non-convergence problems. The effective sample size of the chains ranged from 466 (time discrimination item 1) to 9,000 (time intensity item 1). The naive posterior standard deviation were often close to the time-series standard error. The 10,000 MCMC iterations were sufficiently long and the chains showed convergence after 1,000 iterations.

The {summary} function of {LNIRTQ} provides a similar output as the {LNIRT} function. The posterior mean estimates and (naive) standard deviations of the item parameters are given, in addition to the covariance matrix of the item and person parameters. For reasons of brevity, the correlation matrix of the items and the covariance matrix of the persons are discussed.

The correlation matrix of the items is again given in the order of $a$, $b$, $\phi$, and $\lambda$. The correlation between discrimination and difficulty is around .37, and between time discrimination and intensity around -.53. The difficult items tend to discriminate better in ability than the less difficult items. The high time-intensive items do not discriminate well between speed levels. RTs were hardly affected by increasing working speed for the high time-intensive items. For low time-intensive items this effect on the RTs is much higher. There is a strong correlation between item difficulty and time intensity of .79, which states that the difficult items also took more time to be completed.

\begin{verbatim}
 --- Covariance matrix Items ---

Item Matrix Correlation
    [,1]    [,2]    [,3]    [,4]
[1,] 1.000  0.371  0.050  0.406
[2,] 0.371  1.000 -0.602  0.785
[3,] 0.050 -0.605  1.000 -0.534
[4,] 0.406  0.785 -0.534  1.000

	 --- Covariance matrix Person ---

 Estimated Value
 Covariance Matrix
           Theta Intercept Slope1 Slope2
Theta      0.423     0.114 -0.004 -0.014
Intercept  0.114     0.060  0.000  0.000
Slope1    -0.004     0.000  0.114  0.000
Slope2    -0.014     0.000  0.000  0.063
\end{verbatim}

 The covariance matrix of the random person parameters are given under the label \textit{Theta} (ability), \textit{Intercept} (speed intercept), \textit{Slope1} (speed trend), and \textit{Slope2} (speed quadratic component). The variance in working speed intercepts (i.e. the variance in starting speed of persons) is small (.06). The speed trajectories of the persons show a variance of .11 in trends. The variance in deceleration/acceleration in working speed is around .06. The positive covariance between ability and the speed intercept shows that high-ability persons are more likely to start with a higher speed. The negative covariance between ability and the speed trend shows that the speed trajectories of high-ability persons has a negative trend (decrease in speed), in comparison to low-ability persons. The covariance between ability and the quadratic speed component is around -.014, which means that high-ability persons are more likely to show an acceleration in the negative trend in speed.

For the records with missing values, the imputed data leads to population-average estimates of the random person effects. The ability estimate is around the population average of -.162, and the speed components are around zero. For instance, for the record of 147 the estimated random person effects are (ability, intercept, trend, quadratic):
\begin{verbatim}
R> outchess$Mtheta[147,]
[1] -0.228 -0.018 -0.004 -0.004
\end{verbatim}

\section{Summary and discussion} \label{sec:summary}

Computer-based testing has made it possible to collect more information from respondents to improve our understanding and interpretation of test performances and of the behaviors of respondents. RT, the amount of time a test taker spends on answering an item, has shown to be a very useful source of information. RTs have been used to measure working speed, item time-intensity, or the relationship between speed and accuracy (i.e. speed-accuracy trade-off). The modeling of RTs and its integration in the measurement of ability has led to the development of joint models for RA and RTs. However, there is a lack of software tools that supports a joint model data analysis, which includes recent joint modeling extensions.

The R-package {LNIRT} has been developed with the purpose to provide MCMC estimation methods for (hierarchical) joint models for RA and RT data, while including also new developments in this area. The IRT-based measurement models for RA and RT data have been implemented under different parameterizations to make the program suitable for different users. Imputation methods have been integrated to deal with missing data, while the program can also deal with planned missing by design data. Furthermore,  Bayesian significance tests are included to evaluate person and item fit for RT and RA patterns. The developed person-fit statistics can be used to identify aberrant test takers with respect to their RT pattern, RA pattern or both patterns. Explanatory variables can be used to explain differences between persons and items.

 The LNIRT package has been designed to estimate the parameters of a joint model with multiple link functions (linear, probit) and with non-exponential family distributions. The cross-classified nature of the (random) effects (item and person parameters) further complicates the use of standard (multilevel) software for parameter estimation. Furthermore, Bayesian simulation methods are preferred to handle the required high-dimensional numeric integration for parameter estimation. However, black-box MCMC methods (e.g. JAGS) (1) can be very slow for medium to large data sets and for high-dimensional models (2) cannot integrate identification restrictions in the simulation procedure -- for instance re-scaling latent variables to an identified scale in each MCMC iteration -- (3) necessary identifying restrictions on parameters can lead to complex priors for the model parameters, and (4) the computation of the model-fit tools is complex, since the tools need to be integrated in the model description. The MCMC sampler in the LNIRT package has been designed to collect posterior samples with low autocorrelation and a high effective sample size. This leads to a faster and more efficient algorithm than a black-box MCMC method, which is not designed to optimize the information content of the posterior samples.

The {LNIRT} has some limitations which we hope to address in next releases. Currently, the RA data is limited to binary observations. This is a matter of integrating MCMC schemes for ordinal RA data, which have been discussed by \citet{fox2010}. A more elegant way is to integrate a Gibbs sampler for mixed response types that can deal with RA data with different levels of response types. The joint model is limited to two levels of hierarchy, but extensions to more levels have been discussed by \citet{fox2010} and \citet{kleinentink2008}.

Furthermore, in the differential working speed model test takers can change their speed and the speed components can influence the level of ability. However, the ability component represents a summary measure of accuracy and cannot capture within-individual changes in accuracy. Change in accuracy can be modeled directly with a latent growth model and jointly with a latent growth model for speed. However, binary outcomes contain less information than continuous RTs. This limits the number of growth components that can be estimated and limits the flexibility in describing a pattern of change in accuracy \citep{doi:10.1177/0962280219856375}. The change in accuracy can also be modeled by a hidden markov model \citep{doi:10.1080/00273171.2016.1192983}, where dynamic response behavior is modeled by different item-level states. However, the model complexity increases rapidly when increasing the number of states. Furthermore, the flexibility in modeling change depends on the number of specified states. A more efficient way to model change in accuracy is to adjust a person-specific (or group-specific) discrimination parameter by the level of working speed. Then, to model change in accuracy the contribution of the ability component is adjusted by a discrimination parameter, which level is moderated by working speed. The inclusion of item-specific person-level and person-specific item-level variables to allow the speed-accuracy trade-off to vary between items has been considered by \citet{jintelligence3010021} and \citet{10.3389/fpsyg.2019.01675}. More research is needed to integrate such an approach in LNIRT.

\section*{Acknowledgments}

Ahmet Salih Simsek was supported financially by The Scientific and Technological Research Council of Turkey (TUBITAK) through project number 1059B191800628.

\bibliographystyle{apacite}
\bibliography{refs}

\end{document}